\documentclass[pre,superscriptaddress,notitlepage,nofootinbib, twocolumn,dvipsnames]{revtex4-2}
\pdfoutput=1
\usepackage[left=0.6 in, right=0.6 in, top=0.7 in, bottom = 0.7 in]{geometry}
\usepackage{algorithm}
\usepackage{algorithmic}
\usepackage{graphicx}
\usepackage{amsmath,amssymb,mathtools,bm,physics, tensor, wrapfig}
\usepackage{xcolor}
\definecolor{msugreen}{RGB}{24,69,59}
\usepackage{hyperref}
\hypersetup{
    citecolor=msugreen,            % Use the defined color name
    linkcolor=black,
    urlcolor=blue
}

\renewcommand{\r}{\vb r}

\hypersetup{
    colorlinks=true,
    linkcolor=blue,
    filecolor=magenta,      	
    urlcolor=cyan,
}

\usepackage[colorinlistoftodos]{todonotes}

\begin{document}

\title{Tensor Networks for Liquids in Heterogeneous Systems }

\author{Zachary A.~Johnson}%\footnote{Corresponding author.} \thanks{\email{alicen.carefully@example.com}}
\affiliation{Computational Mathematics, Science and Engineering, Michigan State University, East Lansing, Michigan 48824, USA}
\email{john8248@msu.edu}
\author{Luciano G.~Silvestri}%
\affiliation{Computational Mathematics, Science and Engineering, Michigan State University, East Lansing, Michigan 48824, USA}
\author{Pierson Guthrey}%
\affiliation{Lawrence Livermore National Laboratory, Livermore, CA 94550, USA }
\author{Michael S.~Murillo}%
\affiliation{Computational Mathematics, Science and Engineering, Michigan State University, East Lansing, Michigan 48824, USA}

\begin{abstract}
Many-body correlations in strongly coupled liquids and plasmas are critical for many applications in nanofluids, biology, and fusion-related plasma physics, but their description in fully heterogeneous environments remains challenging due to the high-dimensional equations involved. Recently, tensor network decompositions have emerged as powerful tools for tackling such equations by reducing memory usage and computational complexity. In this paper, we solve for equilibrium density and density-density correlation functions of liquids in confined heterogeneous environments using tensor network methods. We demonstrate that these functions admit high compression when their length-scale dependence is encoded via quantized tensor trains or when their spatial-coordinate dependence is represented in standard tensor-train format, but not with respect to their dependence on distinct particle coordinates.
\end{abstract}

\maketitle

% \tableofcontents

%%%%%%%%%%%%%%%%%%%%%%%%%%%%%%%%%%%%%%%%%%%%%%%%%%%%
%%%%%%%%%%%%%%%%%%%   SECTION  %%%%%%%%%%%%%%%%%%%%%
%%%%%%%%%%%%%%%%%%%%%%%%%%%%%%%%%%%%%%%%%%%%%%%%%%%%

\section{Introduction}
The study of many-body systems near thermal equilibrium is of crucial interest to a variety of fields from physics to biology. The homogeneous limit of classical pair interacting fluids and plasmas is well characterized \cite{hansen2013theory}, but interactions with external potentials, confining geometries, or density gradients significantly complicates the matter and yet is extremely relevant to nanofluidics \cite{Bocquet2010}, plasma physics \cite{zhao2023percolation, bergeson2025experimental}, and soft matter physics \cite{nagel2017experimental}. In this static but inhomogeneous limit, direct simulation with molecular dynamics can be quite useful, but a direct theory based description is needed. Further, these ideas can be extended to near-equilibrium time-dependent problems based on dynamic \cite{marconi1999dynamic, te2020classical} and power \cite{schmidt2022power} density functional theory.

Many-body correlation functions in inhomogeneous systems generically are high dimensional as they depend on relative positions as well as absolute positions. Instead of tackling the inhomogeneous problem directly, one can attempt extensions of the homogeneous result using a local or mixed position approximation \cite{barrett2006some}, formalized in terms of a Taylor expansion \cite{Kremp2004-qq} but are known to break down in highly inhomogeneous settings \cite{lurie2014approach}. The inhomogeneous problem can be formalized as a set of coupled integral-equations using the Ornstein-Zernike equation, extendable to near-equilbrium time-dependent problems \cite{schmidt2022power}. This is a great simplification from the full n-body BBGKY hierarchy of correlation functions, and yet in three-dimensional environments the two-particle correlations are fully six dimensional and thus still represent substantial computational difficulty over the one-dimensional homogeneous version. 

% \textcolor{blue}{A nice way to introduce these ideas is to point out that we fundamentally need $g({\bf r}_1, {\bf r}_2)$, but we generally don't know this function. So, we might try some tricks like: $g({\bf r}_1, {\bf r}_2) \approx g((n({\bf r}_1) + n({\bf r}_2)/2)$ or $g({\bf r}_1, {\bf r}_2) \approx g(({\bf r}_1 + {\bf r}_2)/2)$, etc. In all of these cases (which we should write down more carefully than I did here!) that use a homogenous result to build an approximation, we really throw out the true heterogeneity - we want to {\em actually} compute $g({\bf r}_1, {\bf r}_2)$. There is a very nice paper that lays this out - do you guys know it? What is nice is that it also sayd they don't work well, which motivates this work. 
% }

% Methods for tackling this many-body problem included reduced order modeling based on constructions of orthogonal basis functions can be efficient alternatives \cite{lado2009efficient}, and coarse-grained approaches that optimize

Recently, high dimensional problems exhibiting a curse-of-dimensionality, such as those the above, have been simplified by tensor network formulations as a dimensionality-reduction technique based originally on the tensor train for one-dimensional quantum systems \cite{ORUS2014117}\footnote{Tensor trains are referred to as matrix-product states in the field of many-body quantum mechanics \cite{fannes1992finitely}.}. These tensor networks numerically decompose high dimensional objects into sums and products of lower dimensional objects. Methods for tackling these high-dimensional problems were formalized in \cite{oseledets2011tensor, Oseledets2012} and have been now for a variety of applications including plasma kinetics \cite{Ye_Loureiro_2024}, chemical master equation \cite{Kazeev2014}, turbulence \cite{vonLarcher2019}. These constitute promising alternatives to dimensionality reduction techniques such as coarse-grained theory \cite{noid2023perspective}.

In this paper we solve integral equation theory in a homogeneous and heterogeneous environment using and comparing two tensor train topologies based on either separating the spatial dimensions or the spatial length scales. In section~\ref{sec:Review} we introduce many-body integral equation theory, and the tensor and quantized tensor train formulations. In section~\ref{sec:Applications} we then introduce the Lennard Jones fluid as a testbed for using quantized tensor trains, showing the large compression achievable, and proceed to generalize it to a heterogeneous confined liquid using a standard tensor-train based network, comparing the timing with grid based methods.

%%%%%%%%%%%%%%%%%%%%%%%%%%%%%%%%%%%%%%%%%%%%%%%%%%%%
%%%%%%%%%%%%%%%%%%%   SECTION  %%%%%%%%%%%%%%%%%%%%%
%%%%%%%%%%%%%%%%%%%%%%%%%%%%%%%%%%%%%%%%%%%%%%%%%%%%
\section{Tensor Trains and Equilibrium Problems} \label{sec:Review}
In this section we review and introduce notation for the classical integral equation theory used in this paper, as well as the two different tensor train topologies that we use.

%%%%%%%%%%%%%%%%%%% SUBSECTION %%%%%%%%%%%%%%%%%%%%%
\subsection{Integral Equation Theory for Many-Body Correlations}

The set of problems we consider are that of strongly coupled classical systems in equilibrium at inverse temperature $\beta$ but with density inhomogeneities. Solving for the number density or potential energy formally requires solving a hierarchy of coupled equations in terms of density correlation functions,
\begin{align}
    n^{(p)}(\r_1,\cdots, \r_{p})=\langle \hat{n}(\r_1) \cdots \hat{n}(\r_{p}) \rangle.
\end{align}
Specifying this for all $p$ particles fully determines the system, but is of little practical use. Instead, the first two correlation functions can be used to define most of the thermodynamic properties \cite{hansen2013theory}, which includes the density itself, and the normalized pair correlation function,
\begin{align}
    g^{(2)} (\r_1, \r_2) = \frac{n^{(2)}(\r_1, \r_2)}{n(\r_1)n(\r_2)} - \frac{1}{n(\r_1)} \delta(\r_1-\r_2).\label{eq:gofr_def}
\end{align}
Truncation of the many-body problem at this level can be done in different ways. In kinetics, for which the system of equations takes the form of the BBGKY hierarchy, the Kirkwood approximation \cite{Kirkwood1935} can be used without further information and can be generalized easily to higher order correlation functions. However, in equilibrium cases better closures exist.

% \subsubsection{BBGKY Hierarchy}
% \begin{align} \label{eq:YBG_1}
%     \nabla \left( \ln n(\r)/n_0 + \beta V_{\rm ext}(\r) \right) = \nonumber \\ 
%       -\int d \r' n(\r') g(\r, \r') \nabla \beta u(\r,\r') 
% \end{align}
% These equations are closed by a second set of equations involving pairs of particles, for example,
% $n^{(2)}(\r_1, \r_2)$ is then found through the equilibrium limit of the second level of the BBGKY hierarchy,
% \begin{align}
% \label{eq:YBG_2}
% &\nabla_1 \ln g^{(2)} (\r_1, \r_2) + \nabla_1 \beta u(\r_1, \r_2) =  \nonumber \\
% & g^{(2)} (\r_1, \r_2)\int d\r_3  g^{(2)}(\r_1,\r_3) \:n(\r_3)\nabla_1 \beta u(\r_1,\r_3)  \times \nonumber\\
% &\left[ \frac{g^{(3)}(\vb r_1, \vb r_2, \vb r_3 )}{g^{(2)}(\r_1,\r_3) g^{(2)}(\r_1,\r_2) }  -  1\right] \label{eqn:YBG-2},\\
% \end{align}
% which requires some assumption about the three body density-density-density correlation function, $g^{(3)}$, often taken in the limit of the Kirkwood \cite{Kirkwood1935} approximation
% \begin{align}
%     g^{(3)}(\r_1,\r_2,\r_3)=g^{(2)}(\r_1,\r_2)g^{(2)}(\r_1,\r_3)g^{(2)}(\r_2,\r_3). \label{eq:Kirkwood}
% \end{align}
% The above equations Eq.~\eqref{eq:YBG_1, eq:YBG_2} can be generalized to highger $n$-point correlation functions, and provide a formal means of tackling the $n$-body problem. 

% \subsubsection{Ornstein-Zernicke and Thermodynamic Closures}
In this paper we determine the second order correlation functions using two coupled set of equations. First is the exact Ornstein-Zernike (OZ) equations,
\begin{align}
    h(\r_1, \r_2) = c(\r_1, \r_2) + \int d {\r_3} n(\r_3) c(\r_1,\r_3) h(\r_3, \r_2),
    \label{eq:OZ}
\end{align}
where the direct correlation function $c$ is defined as
\begin{align}
    c(\r_1, \r_2)&=-\beta \frac{\delta^2 \mathcal{F}^{\rm ex}}{\delta n(\r_1)\delta n(\r_2)}  \label{eq:c_def},
\end{align}
in terms of the excess intrinsic free energy $\mathcal{F}^{\rm ex}$. The second equation is where the approximation and truncation of the hierarchy of interactions is done. These include the hypernetted-chain (HNC) approximation, which can be obtained by a second order expansion of the free energy \cite{hansen2013theory}, or its linearized version, Percus-Yevick (PY)
\begin{align}
    % c^{\rm PY}(\r_1, \r_2) &= (1-\exp\left[-\beta u(\r_1, \r_2) \right]) g(\r_1, \r_2) \label{eq: PY_hetero_def}\\
    % g^{\rm HNC}(\r_1, \r_2) &= e^{-\beta u(\r_1, \r_2 ) + h(\r_1, \r_2) - c(\r_1, \r_2)}, \label{eq:HNC_hetero_def}\\
    g^{\rm PY}(\r_1, \r_2) &= e^{-\beta u(\r_1, \r_2)} (1 + h(\r_1, \r_2) - c(\r_1, \r_2)). \label{eq:PY_hetero_def}
\end{align} 
In a mean field limit we have $g\to1$ and $c \to -\beta u(r)$ for inverse temperature $\beta=1/T$, and pair potential $u(r)$. Both closures can be thought of as diagrammatic summations, with HNC incorporating more terms and one can recover PY from HNC in a linear approximation of the quantity $\gamma=h-c$. Despite this, each has their own use, as the PY potential is better for hard-sphere like systems and HNC is better for Coulomb like potentials \cite{hansen2013theory}. In this work we will use only the above PY approximation, discussing its limitations.

With the hierarchy truncated, the density can now be determined using the exact Lovett-Mou-Bluff-Wertheim equation \cite{Lovett1976, wertheim1976correlations},
\begin{align}
    &\nabla \left( \ln n(\r)/n_0 + \beta V_{\rm ext}(\r) \right) = 
      \int d \r' \nabla' n(\r')  c(\r,\r')  \label{eq:Lovett}. 
\end{align}

In either case, we seek to describe second order correlation functions, which in a fully heterogeneous environment are six dimensional and thus exceedingly expensive to work with. Higher order correlation functions further up the hierarchy are even more expensive with less obvious closures. Thus dealing with these equations would be greatly simplified if efficient dimensionality reducing methods could be implemented.  

%%%%%%%%%%%%%%%%%%% SUBSECTION %%%%%%%%%%%%%%%%%%%%%
\subsection{Tensor Train Decompositions}

Tensor networks provide a systematic approach for addressing high-dimensional problems by decomposing functions into products of lower-dimensional tensors. This methodology enables the representation of functions that would otherwise be computationally intractable due to the exponential scaling of traditional grid-based methods. Among the various tensor network architectures, the linear tensor network, or tensor train (TT) decomposition~\cite{oseledets2011tensor}, offers good compression without the computational expense of networks with loops.

The TT decomposition represents a $D$-dimensional function $T(x_1,\ldots, x_D)$ defined over continuous variables $x_i$ through a sequence of three-dimensional tensor cores. Implementation requires discretization of the function over a grid of size $N^D$, where $N$ is the number of points along each continous variable $x_i$, transforming the function into a discrete tensor of dimensions $D$, $T(x_1, \ldots, x_D) \to T_{i_1 \cdots i_D}$. The resulting TT representation takes the form:
\begin{align}
\label{eq:TT_decomp}
% T_{i_1 \cdots i_D} = \sum_{\alpha_1,\ldots,\alpha_{D-1}}^{b_1,\cdots,b_{D-1}} C^{(1)}_{i_1,\alpha_1} C^{(2)}_{\alpha_1,i_2,\alpha_2} \cdots C^{(D)}_{\alpha_{D-1},i_D}
T_{i_1 \cdots i_D} = \sum_{\alpha_1}^{b_1} \ldots \sum_{\alpha_{D-1}}^{b_{D-1}} C^{(1)}_{i_1,\alpha_1} C^{(2)}_{\alpha_1,i_2,\alpha_2} \cdots C^{(D)}_{\alpha_{D-1},i_D}
\end{align}
where each tensor core $\mathbf{C}^{(k)}$ is a three-dimensional array with one physical index $i_k$ corresponding to the discretized variable and two auxiliary indices $\alpha_{k-1}$ and $\alpha_k$ that facilitate the tensor contraction. 
% The auxiliary indices possess dimensions $b_{k-1}$ and $b_k$, respectively, termed bond dimensions, which determine the representational capacity of the decomposition.
\begin{figure}
    \centering
    \includegraphics[width=\linewidth]{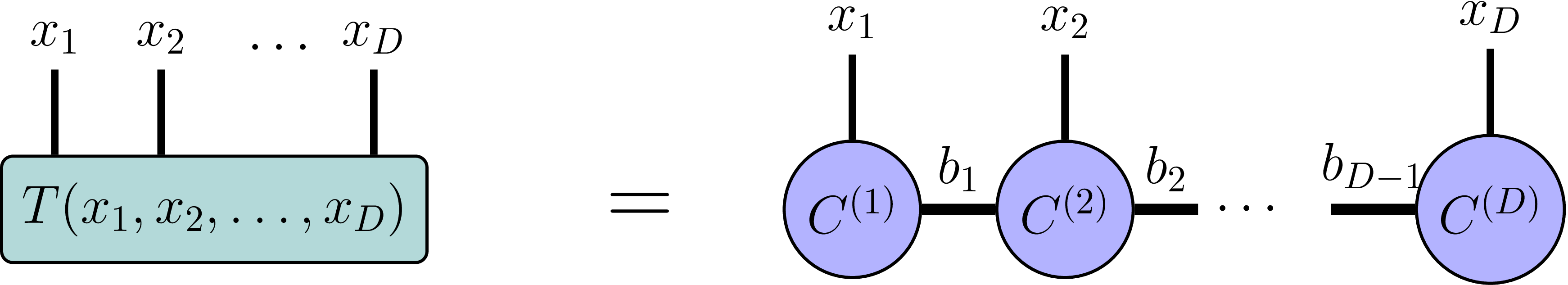}
    \caption{Correspondence between an $D^{\rm th}$ dimension tensor $\bf T$ and it's TT decomposition. The external lines represent the arguments of the initial tensor, and the internal contracted lines represent tensor contractions between cores. The bonds between cores represent summations or contractions; the size or bond dimension $b_i$ of this summation determines the extent of the correlation between adjacent cores \cite{orus2014practical}.}
    \label{fig:TT_diagram}
\end{figure}
The graphical representation of this decomposition, illustrated in Fig.~\ref{fig:TT_diagram}, demonstrates the network structure where uncontracted indices correspond to the original tensor dimensions, while internal bonds represent summations over auxiliary indices. The bond dimensions $b_i$ are the size of these axuilary summations, and quantify the extent of correlations captured between adjacent cores, with larger bond dimensions enabling more accurate representations at increased computational cost~\cite{orus2014practical}.

Construction of the TT decomposition employs iterative partial singular value decomposition (SVD) applied sequentially from left to right, incorporating a tolerance parameter for singular value truncation. For $N$ discretization points per dimension, the computational complexity of the SVD ranges from $\mathcal{O}(N^{D+1})$ to $\mathcal{O}(N^{D+2})$, with the scaling dependent on the degree of compression achieved~\cite{oseledets2011tensor}. Alternative algorithms have been developed to reduce computational overhead, including randomized approaches~\cite{huber2018randomized} and interpolative methods~\cite{OSELEDETS201070, Ritter_2024} that avoid explicit tensor construction.

The efficiency of TT representations depends critically on maintaining small bond dimensions throughout the decomposition. Standard operations on TT formats, including addition and multiplication, have been rigorously defined~\cite{oseledets2011tensor}, though these operations typically increase bond dimensions, necessitating subsequent compression. The TT-SVD rounding procedure addresses this issue by combining QR decomposition with SVD to achieve optimal low-bond dimension approximations with computational scaling of $\mathcal{O}(D N b^3)$, where $b$ denotes the maximum bond dimension before rounding. This rounding step ensures that bond dimensions remain manageable while preserving approximation accuracy within specified tolerances.

\subsection{Quantized tensor train decompositions} \label{sec:QTTs}

Tensor train decompositions achieve optimal compression when functional dependencies between dimensions exhibit weak coupling. For functions where strong interdimensional correlations preclude efficient standard TT compression, the quantized tensor train (QTT) representation~\cite{Khoromskij2011} provides an alternative approach based on hierarchical scale separation~\cite{Shinaoka2023}. In the QTT framework, each tensor core corresponds to a distinct spatial scale, with bonds between cores encoding correlations across different length scales rather than between functional dimensions.
% \begin{figure}
%     \centering
%     \includegraphics[width=\linewidth]{Figures/QTT_diagram_1D-4D.png}
%     \caption{ }
%     \label{fig:QTT_diagram}
% \end{figure}
\begin{figure}
    \centering
    \includegraphics[width=\linewidth]{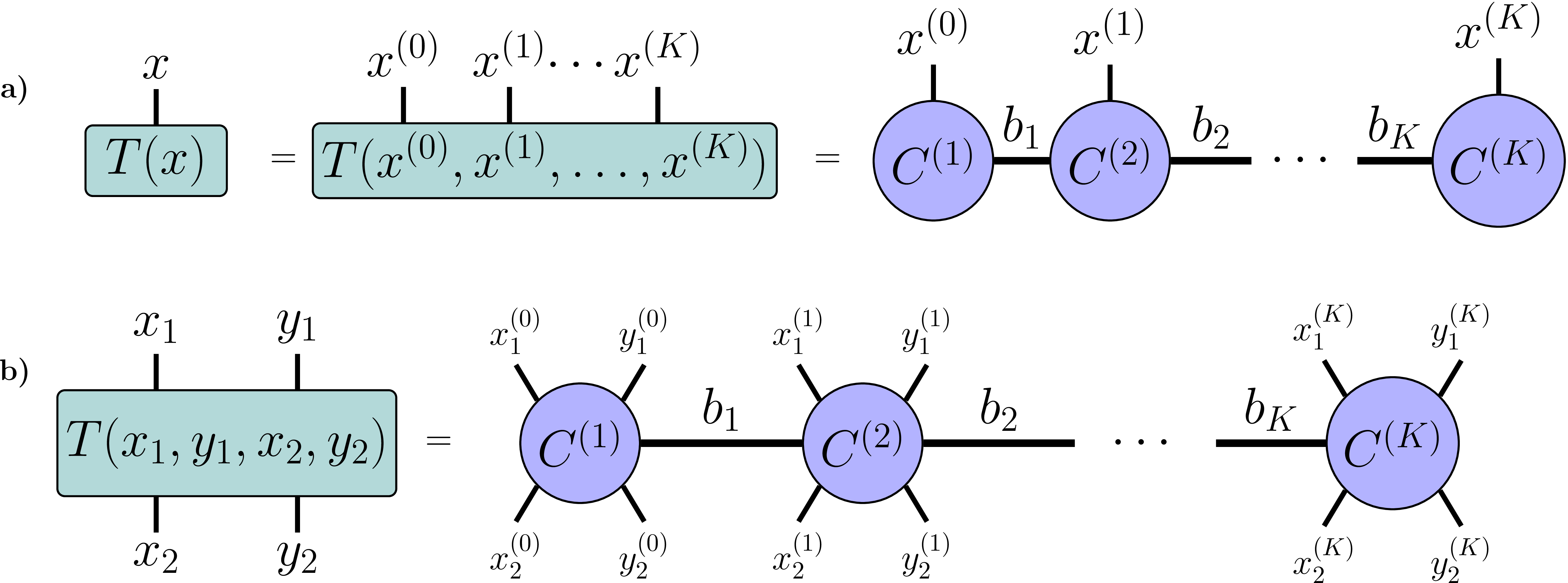}
    \caption{Tensor network diagram where the superscript corresponds to the length-scale shown in Eq.~\eqref{eq:qtt_lengthscales} for the QTT representation of \textit{a)} a one-dimensional function discretized into $2^{K}$ points and \textit{b)} for a pair-correlation style function in two dimensions $f(\vb x_1, \vb x_2)$.}
    \label{fig:QTT_diagram}
\end{figure}
The QTT discretization of a one-dimensional function on a grid of length $N = 2^K$ employs a binary hierarchical decomposition:
\begin{align}
    f(x) = f(x^{(K+1)}/2 + x^{(K)}/2^2 + \cdots + x^{(0)}/2^{K}) \label{eq:qtt_lengthscales}
\end{align}
where each variable $x^{(k)}$ represents the coordinate contribution at the $k$-th scale level. This hierarchical representation, illustrated in Fig.~\ref{fig:QTT_diagram}, enables the decomposition of multidimensional functions through scale-based rather than dimension-based factorization.

For multidimensional functions, each dimension undergoes independent quantization. The resulting tensor cores can be organized in two configurations: grouped arrangement, where all dimensions at each scale level are consolidated into single cores with bonds connecting only different length scales, or interleaved arrangement, where dimensions are distributed across the tensor network. Previous investigations~\cite{PhysRevE.106.035208} demonstrated near-identical compression between these configurations, a result we confirm for the present applications. Given that many TT algorithms exhibit unfavorable scaling with tensor bond dimension, the grouped configuration proves computationally advantageous while maintaining equivalent compression.

To evaluate the relative performance of QTT and standard TT representations, we analyze the compression of a core-smoothed 12-6 Lennard-Jones potential
\begin{align}
     \beta u(\mathbf{r}_1, \mathbf{r}_2) &= 4 \beta^*  \left(\frac{1}{r_{12}^{12}} - \frac{1}{r_{12}^6}\right)  \label{eq:LJ_betau}
\end{align}
expressed in Lennard-Jones units where $\beta^* = \epsilon/T$ and the core-smoothed distance is $r_{12} = (|\mathbf{r}_1 - \mathbf{r}_2|^4 + 0.4^4)^{1/4}/\sigma$, with $\sigma$ representing the characteristic Lennard-Jones length scale. Core smoothing prevents numerical instabilities associated with sharp cutoffs while preserving the essential physics in the limit of vanishing smoothing parameter.

\begin{figure}[ht]
    \centering
    \includegraphics[width=\linewidth]{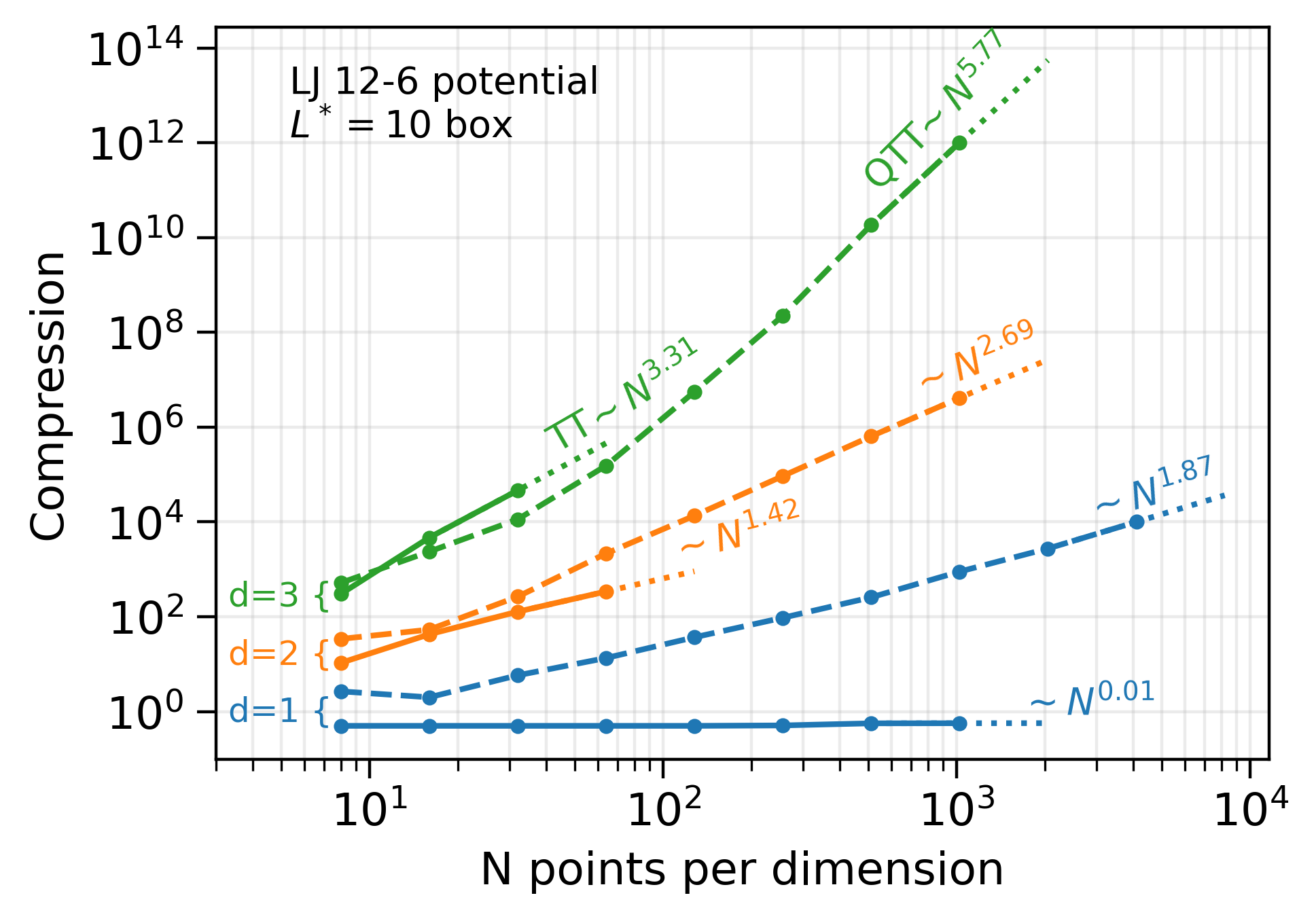}
    \caption{Storage compression comparison for the 12-6 Lennard-Jones potential Eq.~\eqref{eq:LJ_betau} represented on $2d$-dimensional grids of length $L = 10$ along each dimensions in Lennard-Jones units. QTT (solid lines) and TT (dashed lines) representations assume singular value cutoffs of $10^{-4}$. Asymptotic scaling behaviors are indicated by dotted lines. }
    \label{fig:compression}
\end{figure}

Figure~\ref{fig:compression} presents compression analysis for this pair potential across one to three spatial dimensions, corresponding to a two to six dimensional pair potential. 
% \LGS{The TT representation employs 2 cores for each spatial dimension, meaning that in the 1D case $\beta u(\vb r_1, \vb r_2) = \beta u(x_1, x_2)$ is represented by 2 cores, in the 2D case $\beta u(\vb r_1, \vb r_2) = \beta u(x_1, x_2, y_1, y_2)$ is represented by 4 cores} 
The TT representation employs $2d$ cores for $d$-dimensional systems, while the QTT with grouped arrangement used in this work utilizes $K=\log_2 N$ cores, see Fig.~\ref{fig:QTT_diagram}b. TT decompositions are constructed through explicit grid discretization followed by SVD truncated at $\epsilon$ of the full Frobenius norm of the singular values; this procedure is expensive and so we only go to limited grid resolutions. Instead for QTT representations we employ tensor-cross interpolation (TCI)~\cite{xfac}, which enables finer grid resolution at the expense of potential sampling errors. We define the error of TCI using the relative norm
\begin{align}
    {\rm Err}(f_1, f_2) = 2\frac{||f_1 - f_2||}{||f_1|| + ||f_2||}, \label{eq:err_def}
\end{align} 
where the norm is the Frobenius norm. Empirical analysis reveals TCI limitations in achieving relative errors below $10^{-4}$ for these functions, requiring careful initialization with multiple strategically chosen pivot points. Additionally, TCI exhibits poor convergence for functions with discontinuities or kinds, necessitating the core-smoothed potential formulation.
 
The QTT storage scaling asymptotically approaches $\mathcal{O}(b^2 2^{2d}\log N)$ for function-dependent bond dimension $b$, achieving this scaling once all relevant length scales are captured. In contrast, TT compression demonstrates efficiency only for $d>1$. We find that bonds connecting different particle coordinates (e.g., $x_1, x_2$) exhibit minimal compressibility, indicated by the flat solid blue line in Fig.~\ref{fig:compression}, while bonds between spatial dimensions of individual particles (e.g., $x_1, y_1$) achieve substantial compression. This behavior reflects the fundamental challenge TT representations face for functions of the form $f(|\mathbf{x}_1 - \mathbf{x}_2|)$, characteristic of many-body correlation functions. This contrasts markedly with functions such as $f(x_1 + x_2)$ or $f(x_1^2 + x_2^2)$, which can achieve unit bond dimensions. 
%\LGS{In order to strengthen the paper, I feel that this paragraph needs a plot showing the correlations between $x_1,x_2$ vs $x_1, y_1$. This discussion is referenced later when we solve the inhomogeneous problem. This could be the place to put a plot similar to that in Figure 5.}

The asymptotic TT compression scales approximately as $N^{2d}/(b^2N^2)$, where the bond dimension $b$ between spatial dimensions increases slowly with $N$. Consequently, optimal TT representations of $p$-particle correlation functions in $d$ dimensions are limited to length $d$. This constraint implies that standard TT approaches are insufficient for addressing higher-order terms in the BBGKY hierarchy without additional methodological developments.

The qualitative compression characteristics illustrated in Fig.~\ref{fig:compression} remain consistent across various pair potentials and physical functions exhibiting dominant dependence on $|\mathbf{r}_1 - \mathbf{r}_2|^2$, including correlation functions $h$ and $c$. At low dimensions or for applications requiring high resolution or extensive grids, QTT representations demonstrate superior compression, translating to faster operations. However, at higher dimensions with moderate grid requirements, TT representations become advantageous, as evidenced by the crossover points between QTT and TT performance curves where such intersections exist.
\section{Applications} \label{sec:Applications}

We consider two geometries in which a Lennard Jones fluid defined by Eq.~\eqref{eq:LJ_betau}, is at equilibrium. First, a three-dimensional case with no external potential, representing a homogeneous fluid for which we use the QTT representation, and the second is a liquid confined between two parallel plates for which we use the TT representation.
 
\subsection{Quantized tensor train for homogeneous fluids} \label{sec:QTT_Homo}
For homogeneous fluids the density is a single number, and the pair correlations are one dimensional. Such a limit is particularly well used in the theory of plasmas where one can use quantum-accurate pair potentials with the OZ and HNC equations to obtain plasma properties \cite{stanek2021efficacy, starrett2014simple}, for non-equilibrium systems \cite{2411.02363, hou2017average} and for accurate initialization of expensive MD simulations \cite{2505.04362}.

In the uniform or homogeneous limit, the OZ equation~\eqref{eq:OZ} is stated most simply in Fourier Space using as   
\begin{align}
    \tilde{h}(k) = \tilde{c}(k) + n \tilde{c}(k) \tilde{h}(k), \label{eq:Homo_OZ}
\end{align}
where $n$ is the uniform density of the species, and $\tilde{h}$ is the FT of the pair correlation function $h(\vb r_1, \vb r_2) = h(r_{12})$ for $r_{12}=|\vb r_1 - \vb r_2|$. The Fourier transform hinges on the development of the superfast Fourier transform (SFT), see \cite{Dolgov2012} and Appendix~\ref{app:FT_dst}.

It is convenient to define a common interaction expansion parameter, the Mayer function,
\begin{align}
    f_M(r) = e^{-\beta u} - 1.
\end{align}
This serves as a good initial guess in the QTT representation for both $h$ and $c$. We also use it to rewrite the PY approximation to improve convergence properties. Letting $\gamma = h-c$ and its FT $\tilde{\gamma}$, we rewrite Eq.~\eqref{eq:PY_hetero_def} as
\begin{align}
        h^{\rm PY}(r) &= f_M(r) + (1+f_M(r))\gamma(r), \label{eq:Homo_PY_h}\\
        c^{\rm PY}(r) &= f_M(r)\left (1 +  \gamma(r) \right ). \label{eq:Homo_PY_c}
\end{align} 
% and in the HNC approximation
% \begin{align}
%         h^{\rm HNC}(r) &= -1 + (1+f(r))\exp[\gamma(r)], \label{eq:Homo_HNC_h}\\
%         c^{\rm HNC}(r) &= (1+f(r))\exp[ \gamma(r)] - 1 - \gamma(r). \label{eq:Homo_HNC_c}
% \end{align}

We solve eqs.~\eqref{eq:Homo_OZ}--\eqref{eq:Homo_PY_c} self-consistently---at each step we obtain a new $c$ and $h$, which we mix with the previous iteration using $h_{i+1} = \alpha h^{\rm PY}(\gamma_i(r)) + (\alpha-1) h_i$ for mixing parameter $\alpha=0.1$, and likewise for $c$. We round at every multiplication and addition step; bonds increase by doubling or squaring with each addition or multiplication so for a rounding cost of $\sim b^3$, it is crucial to round before bonds grow substantially. During the algorithm it is never necessary to unfold the QTT, and thus nearly arbitrarily high resolution can be obtained. One can truncate the QTT at a given low spatial resolution and carry out the leftover matrix multiplications to obtain unfolded representations for examination and plotting at low cost. 

\begin{figure}[ht] % Adjust [ht] for placement (t=top, b=bottom, h=here)
    \centering % Center the entire figure
    \includegraphics[width=\linewidth]{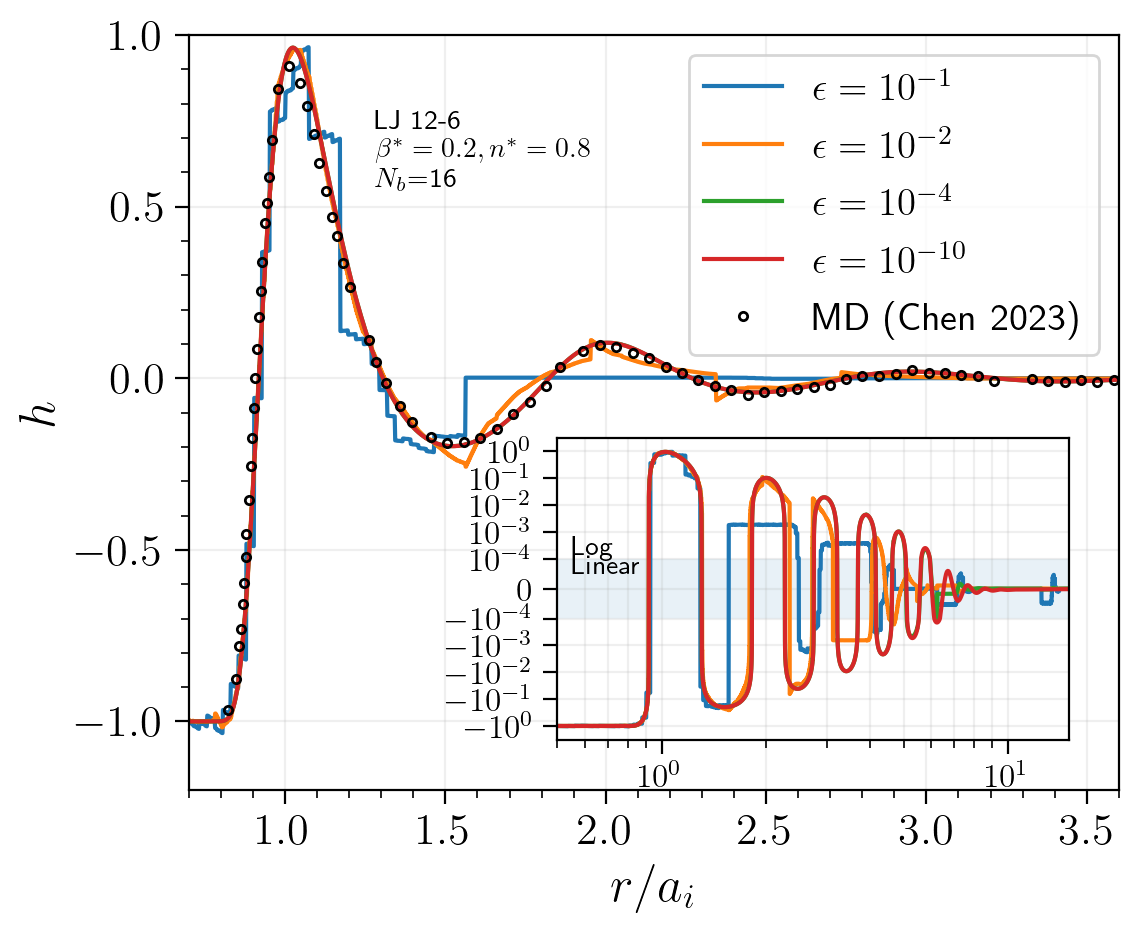} % Or .jpg, .pdf, etc.
    \caption{Radial distribution function for a 12-6 LJ fluid from Eq.~\eqref{eq:LJ_betau} and parameters in the inset. We show QTT solutions with the four SVD tolerances shown. We additionally compare against MD data from \cite{chen2023connection} finding errors, Eq.~\eqref{eq:err_def}, of  0.176083 (blue, $\epsilon=10^{-1}$), 0.116417 (orange, $\epsilon=10^{-2}$), 0.115816 (green, $\epsilon=10^{-4}$), 0.115814 (red, $\epsilon=10^{-10}$)). The inset plot is a mixed log, linear plot (symlog) of $h$ to show the long-distance oscillations.}% Optional: caption for the first subplot
    \label{fig:Homo_RDF_plot} % Optional: label for the first subplot
\end{figure}

In Fig.~\ref{fig:Homo_RDF_plot} we show converged solutions of the pair correlation function, $h$ where the TTs are always rounded to TT-SVD cutoffs of $\epsilon=0.1$ (blue) to $\epsilon=10^{-10}$ (red). We solve the same set of equations above using a standard gridded representation at a lower resolution, and find the set SVD tolerance corresponds to a relative error with the explicit grid of within a factor of three. It also controls the tolerance achievable in the self-consistent calculation, with rough agreement again depending on the function behaviour. We then compare our results with MD results from \cite{chen2023connection} and find close agreement, see Figure caption for the explicit errors. 

We see all QTT results agree quite well with the MD, with error essentially unchanged beyond a tolerance of $10^{-1}$; the Frobenius norm error metric itself is biased to areas with large slopes, which where most error occurs for all singular value cutoffs shown. The near invariance of error with respect to singular value cutoff shows the PY model error itself, however at low QTT bond dimension the function is not very smooth, and does a particularly poor job at describing long distance correlations as seen in the symlog inset plot where in particular we can see low-amplitude oscillations in the $\epsilon=0.1$ QTT at arbitrarily large distances.

% \begin{figure*}[ht] % Adjust [ht] for placement (t=top, b=bottom, h=here)
%     \centering % Center the entire figure
%     \begin{subfigure}[b]{0.45\textwidth}
%         \centering
%         \includegraphics[width=\linewidth]{Figures/QTT_Homo_LJ_RDF_plot.png} % Or .jpg, .pdf, etc.
%     \end{subfigure}
%       \hfill
%     \begin{subfigure}[b]{0.45\textwidth}
%         \centering
%         \includegraphics[width=\linewidth]{Figures/QTT_Homo_LJ_RDF_bond_plot.png} % Or .jpg, .pdf, etc.
%     \end{subfigure}
%     \caption{Radial distribution function for a 12-6 LJ fluid with $\beta^*=0.2$ (blue) and $\beta^*=1.0$ (orange). We have the answer using entirely QTT representation (solid) in the PY approximation, compared with a standard gridded representation for the HNC approximation (dashed). We note the grid and QTT representation recover the same answer when the same closure is used. We additionally compare against MD data from \cite{chen2023connection}.  \ZJ{Add grid PY, or rethink purpose. } } % Optional: caption for the first subplot
%     \label{fig:Homo_RDF_plot} % Optional: label for the first subplot
% \end{figure*}

\begin{figure}
    \includegraphics[width=\linewidth]{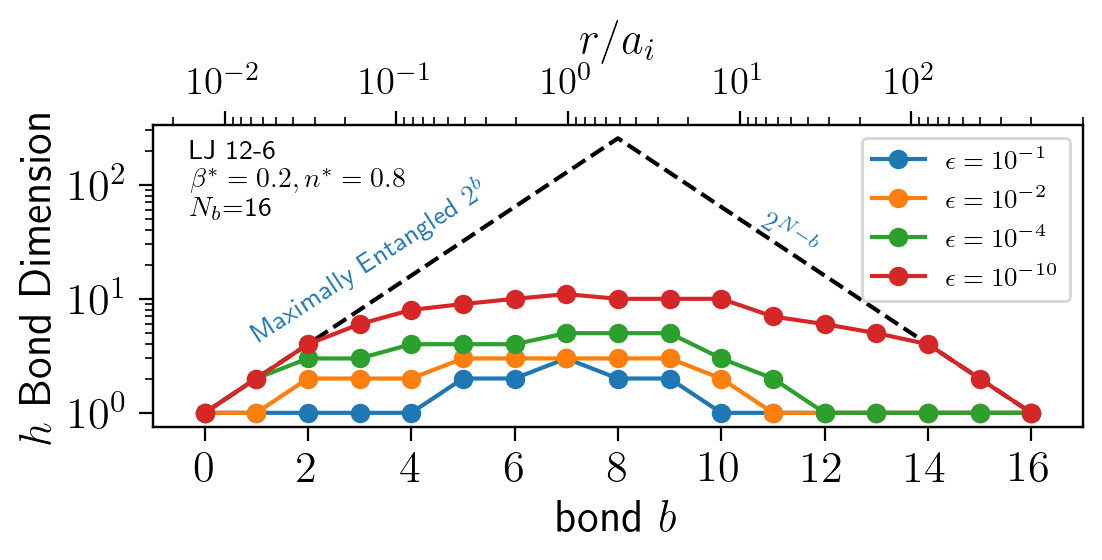} % Or .jpg, .pdf, etc.
    \caption{Plot of the bond dimension of the QTT cores between different length scales of the pair correlation functions in Fig.~\ref{fig:Homo_RDF_plot} where from bottom to top we assume SVD tolerances of ($10^{-1}$, $10^{-2}$, $10^{-4}$, $10^{-10}$). The dashed line displays the theoretical maximum which would correspond to roughly the original full grid representation. The upper axis corresponds to the length scale probed at each bond, as in Eq.~\eqref{eq:qtt_lengthscales}. Representing these QTT's takes, from bottom to top, $(68, 132, 316, 1672)$ floats vs a grid size of $2^{16}=65,536.$ }   % Optional: caption for the second subplot
    \label{fig:Homo_RDF_bond_plot} % Optional: label for the entire figure
\end{figure}

Corresponding to the solutions in Fig.~\ref{fig:Homo_RDF_plot}, we can see the QTT compression as a function of length-scale (core) via a plot of the bond-dimension, displayed in Fig.~\ref{fig:Homo_RDF_bond_plot}. The lower x axis shows the bond on either side of each core, and the upper axis displays the length-scale defined in Eq.~\eqref{eq:qtt_lengthscales}. 
For the cheapest QTT, with a cutoff of $0.1$ for the singular values, we can see most of the information is required around a length-scale of $1$ ion-sphere radii, where most of the complex spatial dependence is. Lowering the SVD cutoff, we first see the bond dimension increasing at low length-scales to smooth out the variations for small length-scales. Perfectly getting the long-range behaviour increases the rightward bond dimensions substantially, however, we can see in Fig.~\ref{fig:Homo_RDF_plot} that the green line approximates the function well within $\epsilon=10^{-4}$ with only unit bond dimensions.

The number of floats needed to represent this function, listed in the caption in Fig.~\ref{fig:Homo_RDF_bond_plot} is on the order of hundreds, despite representing $N=65,536$ grid points. This grid, however, is much denser than is needed in practice; a better comparison is to a grid defined by the minimum size that covers the functions non-zero extent, looks smooth to the eye, and converges correctly. A grid of length $10$ $a_i$  and $N\sim500$ points satisfies this criterion, which is roughly comparable to the QTT memory usage with $\epsilon=10^{-4}$.  

In addition to the compression, the timing of the grid based vs QTT algorithms depends sensitively on the grid discretization and SVD truncation used. The most expensive operations for the QTT are the rounding after multiplication with complexity $\mathcal{O}(2 b^6)$, and the SFT transform with complexity $\mathcal{O}\left( \log^2(N) b^3 \right)$. Generally we find the SFT comparable to the multiplication and rounding step for $\epsilon=10^{-10}$, but as the error is allowed to increase, the bond dimensions drop and the SVD step becomes negligible. The timing of the grid based method is based largely on the FFT timing of $\mathcal{O}(N \log(N))$, and we find comparable QTT and grid timings only for grids bigger than $N\gtrsim10^7$. This implies that for 1D problems such as this, QTT's are only faster than grid based methods if the functions involved require extremely dense or large grids. In higher, $d$, dimensions, the SFT scales only linearly with $d$, whereas the FFT scales as $N^d$, so for FFT applications we expect QTT's to be much faster in higher dimensions.

%%%%%%%%%%%%%%%%%%% SUBSECTION %%%%%%%%%%%%%%%%%%%%%

\subsection{Tensor trains for liquids confined between parallel plates} \label{sec:TT_1D_asymmetry}
In this section we consider systems with an asymmetry in one direction, which well approximates many systems including liquids near a wall, Yukawa plasmas in a confining potential as well as shock fronts. In this limit the density is a one-dimensional function and the two-body correlation functions can be reduced to be three dimensional, see Appendix~\ref{app:dof}. 
The LJ potential is the same as in the previous section and we have two parallel plates with the external potential,
\begin{align}
    \beta V_{\rm ext}(x) &= 2 \pi \beta^*  \left( \frac{2}{5}\frac{1}{{x}^{10}} - \frac{1}{{x}^4} - \frac{1}{3}\frac{1}{\delta x ( x + 0.61 \delta x)^3} \right) \label{eq:LJ_Vext},
 \end{align} 
from walls at $x=0$ and $x=L$, where we again assume LJ units. This external potential is designed to mimic the potential from a series of LJ walls spaced by $\delta x=1/\sqrt{2}$\cite{Snook1980}. 
 
We define a three-dimensional grid $N_x \times N_x \times N_\rho$ on which the correlation functions $c(x_1, x_2, \rho)$ where $\rho=|\rho_{12}|$ is the distance in the radial direction. It is simplifying to work in Fourier space in the radial direction, parallel to the wall surface, \cite{plischke1986density}, so in Fourier space we have $\tilde{c}(x_1, x_2, k)$ where $N_\rho=N_k \approx N_x$, see Appendix~\ref{app:FT_Hankel} for details. The number of grid points is only approximately equal because we remove axial grid points in the small exclusion zone next to the walls. 
 
\begin{figure}
    \centering    \includegraphics[width=0.9\linewidth]{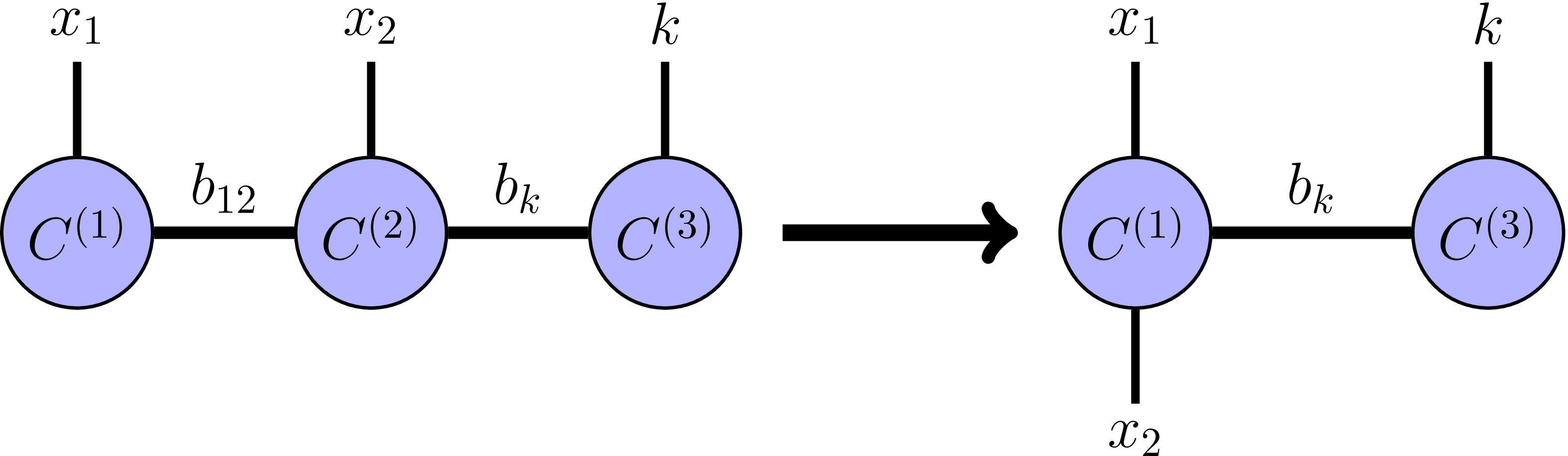}
    \caption{On the left is the TT version where each dimension corresponds to a core. For functions dependent strongly on $|x_1-x_2|$, the bond dimension $b_{12}\sim N$ for $N$ the size of the grid in the $x$ direction, see discussion around Fig.~\ref{fig:compression}. Characteristic TT algorithms scale to high powers of the bond dimension, making the rightward implementation much cheaper in this high bond dimension example.}
    \label{fig:1D_TT_diagram}
\end{figure}

 We then form a TT rather than a QTT in the form shown on the right in Fig.~\ref{fig:1D_TT_diagram}. We choose to use this architecture because in this case we only FT one direction, which cannot be done using a QTT format, although in principle one could consider mixed QTT-TT formats, which we did not do. We form our initial guess as before with $f_M$, except that we use the TT-SVD algorithm~\cite{oseledets2011tensor} to form the initial TT. This is expensive relative to other individual operations, but must only be done once to initialize the calculation. 
 % within a specified tolerance of $\epsilon=10^{-4}$. 
We employ the TT topology shown on the right of Fig.~\ref{fig:1D_TT_diagram} rather than the left because as previously discussed surrounding Fig.~\ref{fig:compression}, the bond dimension between $x_1, x_2$ is near maximal, or $b_{12} \sim N_x$, whereas the functions we consider are highly compressible between the transverse and longitudinal directions. One could still form a TT representation in the form of the left of Fig.~\ref{fig:1D_TT_diagram} for no real loss in compression, however it makes multiplication and rounding substantially more expensive. With the notation $b_{12}$ referring to the bond dimension between $x_1$ and $x_2$ cores, and $b_k$ for the bond to the last core, Hadamard (point-by-point) multiplication scales as $\mathcal{O}(N_x b_{12}^2 b_k^2)$, with subsequent rounding scaling as $\mathcal{O}(N_x b_{12}^2 b_k^4)\sim\mathcal{O}(N_x^3 b_k^4)$ compared to the grid-based scaling of $\mathcal{O}(N_x^2 N_k)$. The integration in Eq.~\eqref{eq:OZ_1D} is even worse since one must naively insert a new core corresponding to $x_3$. 

Instead, if we put both $x_1$ and $x_2$ in the same core, then the OZ equation, 
\begin{align}
    h(x_1, x_2, k) = c(x_1, x_2, k) + \int dx_3 n(x_2) c(x_1,x_3,k) h(x_3, x_2, k). \label{eq:OZ_1D}
\end{align}
becomes matrix multiplication over the longitudinal $x_{12}$ core weighted by $n$ with cost $\mathcal{O}(N_x^3 b^2 + N_k b^2 )$ vs a grid cost of $\mathcal{O}(N_x^3 N_k)$. The bond dimension squares during multiplication and is rounded by the TT-SVD algorithm~\cite{oseledets2011tensor} for cost $\mathcal{O}(N_x^2 b_k^4)$. Thus, the TT calculation should beat the grid-based when $b_k^2\lesssim N_x$, assuming $N_x\sim N_k$.

The second order correlation functions are then closed by the PY approximation,
\begin{align}
        h^{\rm PY}(x_1, x_2,\rho_{12}) &= f_M(x_1, x_2,\rho_{12}) +\nonumber\\
         & (1+f_M(x_1, x_2,\rho_{12}))\gamma(x_1, x_2,\rho_{12}), \label{eq:PY_1D_h}\\
        c^{\rm PY}(x_1, x_2,\rho_{12}) &= f_M(x_1, x_2,\rho_{12})\left (1 +  \gamma(x_1, x_2,\rho_{12}) \right ). \label{eq:PY_1D_c}
\end{align}
The resulting $c$ is then inserted into eq.~\eqref{eq:Lovett},
\begin{align}
    \nabla_{x_1} \ln n(x_1) + \nabla_{x_1}  \beta V_{\rm ext}(x_1) &= \int d x_2 \nabla_{x_2} n(x_2) \tilde{c}(x_1, x_2, 0).\label{eq:Lovett_1D}
\end{align}
Equations.~(\ref{eq:OZ_1D}--\ref{eq:Lovett_1D}) are solved self-consistently by mixing with a coefficient $\alpha=0.1$, but the density is explicitly minimized for a given $\tilde{c}(x_1,x_2,0)$  using a least squares algorithm, and this solution is then also mixed with the previous iteration using the same mixing fraction. 

\begin{figure}
    \centering
    \includegraphics[width=\linewidth]{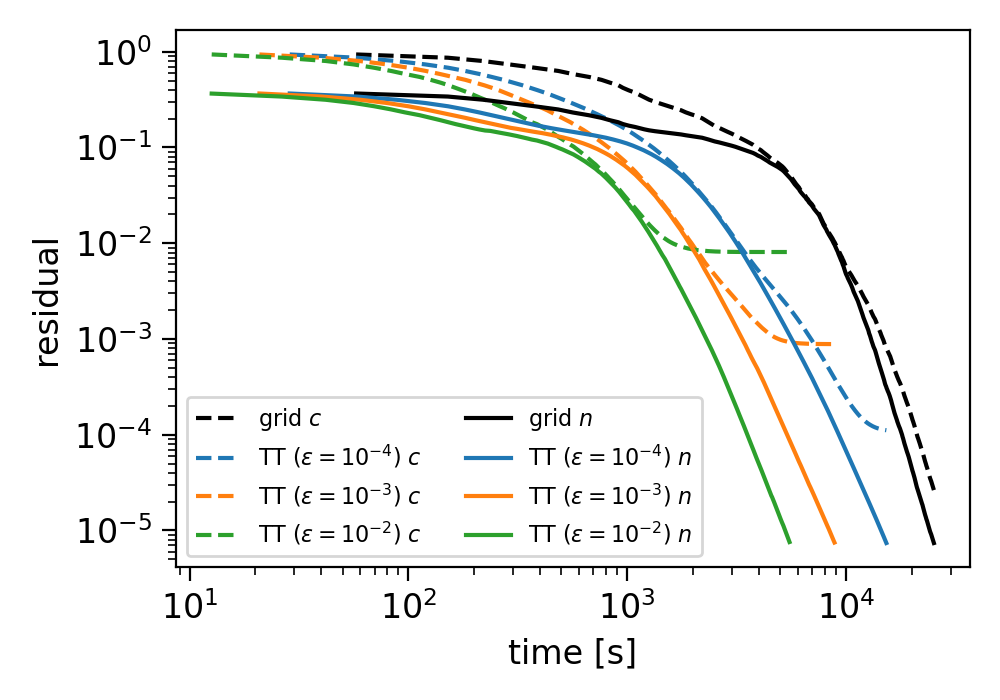}
    \caption{Timing of the pure grid based (black lines) and TT based methods (colored lines) for convergence of the direct correlation function (dashed) and number density (solid) for the LJ fluid between LJ plates problem with $L^*=3$, $\beta^*=0.02$, $n^*=1$. Each color has a different singular value cutoff for the TT's bonds, with final bond dimensions of $b_k= (3, 6, 11)$ for cutoffs $\epsilon=(10^{-2}, 10^{-3}, 10^{-4})$. }
    \label{fig:TT_1D_timing}
\end{figure}

We monitor the convergence of the calculation by tracking the residual, Eq.~\eqref{eq:err_def}, between successive iterations before mixing. Figure~\ref{fig:TT_1D_timing} shows this convergence behavior for both $c$ and $n$ as a function of time for different TT SVD cutoffs. We can see that the singular value cutoff determines roughly the asymptotic accuracy of the self-consistent calculation, though we note that the residual for $h$, not shown, is consistently higher than for $c$. The bond dimensions for all cases shown in this plot obey $b_k^2 < N_x$, and thus we find the TT algorithm to be faster. Note that since the algorithm is fundamentally unchanged between the grid and TT approaches, re-plotting Fig.~\ref{fig:TT_1D_timing} as a function of iteration yields overlapping lines until the iteration where the given TT approximation error itself causes the residual to saturate.

\begin{figure*}[t!]
    \centering
    \includegraphics[width=\linewidth]{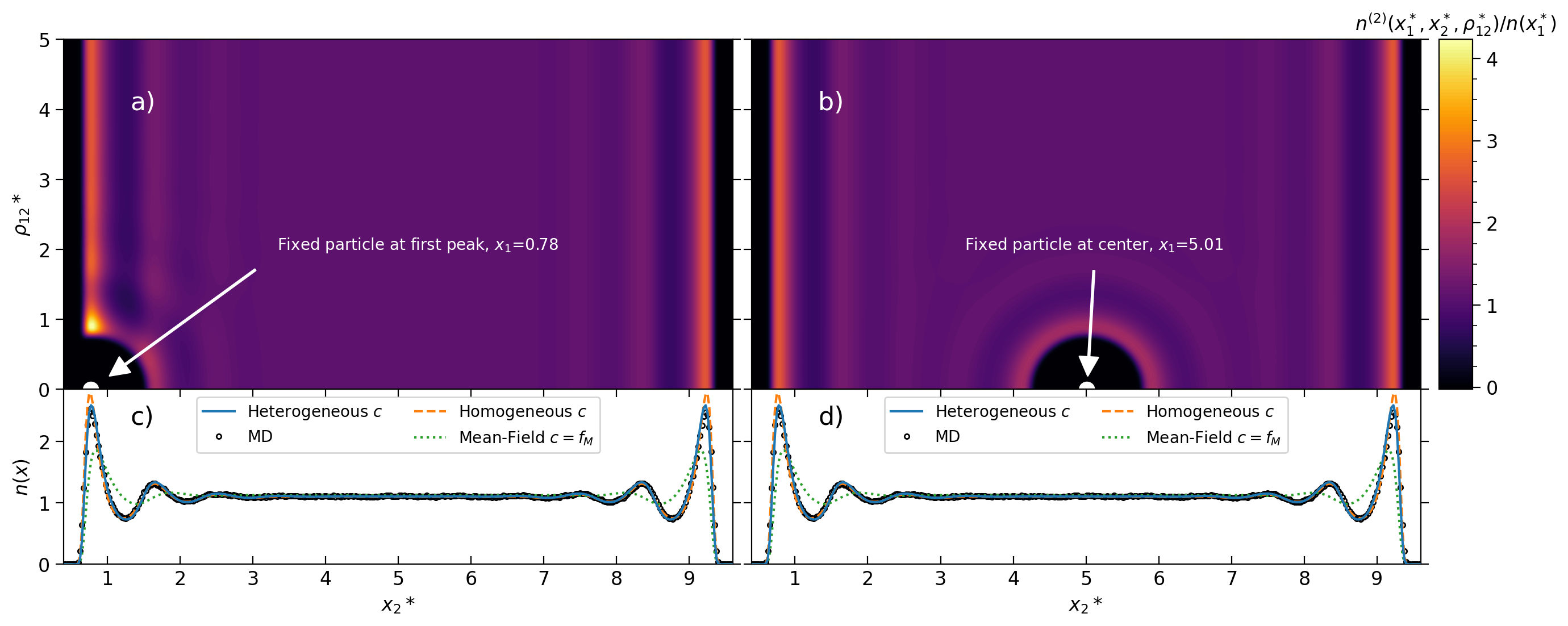}
    \caption{ Solution to a 12-6 LJ fluid with $\beta^*=0.02$, $n_0^*=1$ between same coefficient 10-4-3 LJ parallel walls 10 LJ units apart. Shown are contour plots of the density-density correlation $n^{(2)}$ normalized by the density at the location of the second particle, assuming the first particle is fixed at a) the leftmost peak, or b) in the center between the plates, thus this is a contour map of the probability of a second particle appearing at a given $x_2$ for a given first particle placement. In c) and d) we have the density of the LJ fluid for our full calculation with TT's (blue solid), the solution to the Lovett equation, but with different approximations. Homogeneous implies $c$ is from a fully homogeneous calculation at the specified density as in Sec.~\ref{sec:QTT_Homo}, and the green-dotted line represents the Lovett solution assuming our initial guess for the direct correlation function, $c=f_M$.    }
    \label{fig:TT_1D_contour_L10}
\end{figure*}

In Fig.~\ref{fig:TT_1D_contour_L10} we compare the fully self-consistent density with heterogeneous correlation functions for the case of walls a distance of $10$ LJ units apart. In the lower panels $c), d)$ we show the density of the fully heterogeneous computation (blue solid) compared against the case with the homogeneous solution to the correlation functions (orange dashed), the approximately mean-field initial guess used in the self-consistent process, $c=h=f_M$ (green dotted), as well as explicit simulation results from our MD simulations with LAMMPS \cite{LAMMPS}. Close to the walls we can see large density oscillations with both homogeneous and heterogeneous calculations quite close to the MD result except for the first peak where the heterogeneous answer more closely agrees with MD. 

Further into the bulk of the liquid the density oscillations die out. In  Fig.~\ref{fig:TT_1D_contour_L10} upper panels we see a slice of the three-dimensional correlations functions where a) one particle is fixed at the highest density peak, and b) where one particle is fixed in the center between the plates. The color then shown is the density-density correlation function which in this slice is directly proportional to the probability of finding a second particle at a given location, thus it is zero adjacent to the wall and large at the first density oscillation. It also shows the isotropy at short distances in the bulk case in panel b). 

\begin{figure*}[t!]
        \centering
    \includegraphics[width=\linewidth]{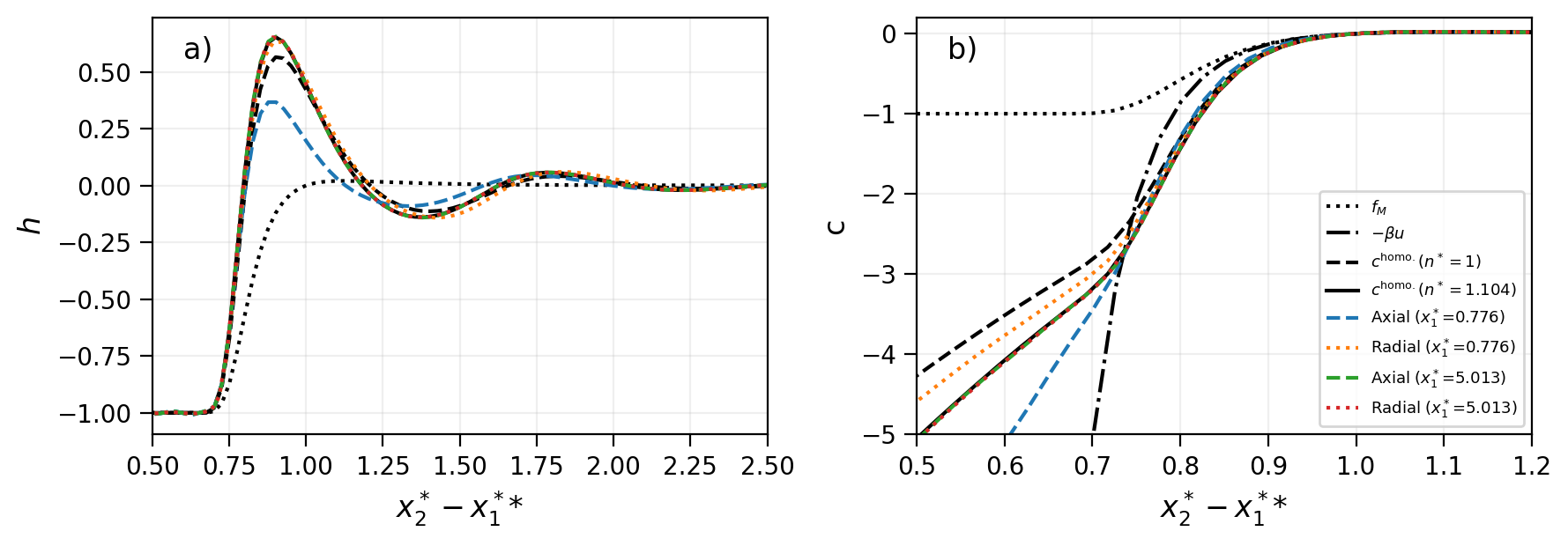}
    \caption{Plotted are 1D slices of a) the pair correlation function, $h(x_1, x_2, \rho )$ with $x_1$ fixed at the first peak and inwards axially (blue dashed), radially (orange dotted), or at the center between the two plates and axially out (green) or radially (red dotted) for the same configuration as Fig.~\ref{fig:TT_1D_contour_L10}. We compare to the homogeneous solution with $n=1$, (black dashed) and the central density, $n=1.104$, (black solid) and the solver initial guess given by the Mayer function (black dotted). In b) we have the direct correlation function defined with the same slices, and additionally the mean-field approximation $-\beta u$ (black dash-dotted). We see in both plots that the middle of the plates the homogeneous approximation is quite good, but near the walls there is anisotropy.}
    \label{fig:1D_TT_ch_lines}
\end{figure*}

We further examine this by comparing one-dimensional slices of the correlation functions $h$ and $c$ in Fig~\ref{fig:1D_TT_ch_lines} wherein one particle corresponding to $x_1$ is fixed at either the first peak or centered between the plates. Recall that $h$ is related to the contour plot via $h(\vb r_1,\vb r_2) = n^{(2)}(\vb r_1, \vb r_2 )n^{-1}(\vb r_1) n^{-1}(\vb r_2) - 1$ so the slices shown of $h$ in panel a) are related to but do not quite correspond to the second particle placement probability\footnote{Technically $n^{(2)}$ contains an additional delta function that we ignore here for plotting purposes.}. In both panels we have the initial guess for the self-consistent calculation, the Mayer function (black dotted), as well as two homogeneous solutions corresponding to the overall density (black dashed) as well as the actual density at the center point between the plates (black dashed). The homogeneous case with the increased density is the one that agrees best with the correlation functions in the center of the plates (green dashed, red dotted). The correlation function slice starting from the first peak but moving radially outward (orange dotted) is also substantially similar, showing that directions perpendicular to the asymmetry are less sensitive to it. The correlation function slices starting from the first peak that go longitudinally to the middle (blue dashed) is substantially different, with reduced peak height and radially (orange dotted).

\begin{figure*}[t!]
    \centering
    \includegraphics[width=0.8\linewidth]{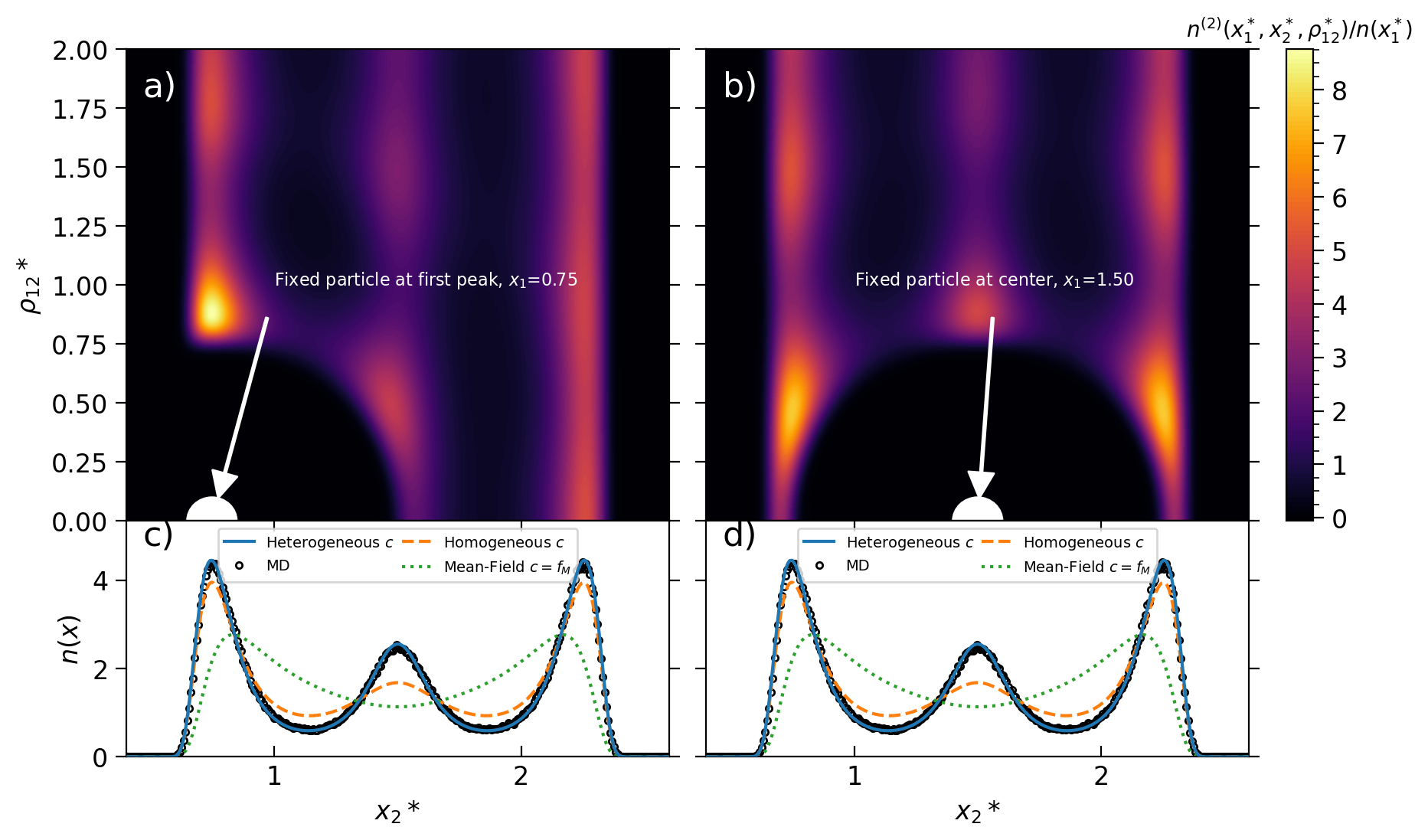}
    \caption{ Identical to Fig.~\ref{fig:TT_1D_contour_L10} but with the plates a distance of only $L=3$ away from each other. In this case we can see a tri-layer crystal start to form and the homogeneous answer is no longer a good approximation anywhere. }
    \label{fig:TT_1D_contour_L3}
\end{figure*}

In the case shown in Fig.~\ref{fig:TT_1D_contour_L3} the liquid is confined by plates that are now much closer, at $L=3$ LJ units separation. In this case we can see from lower panels c) d) that the homogeneous approximation is no longer accurate anywhere and yet the fully self-consistent equation is remarkably close to the MD result. This is true for all three TT SVD cutoffs employed; since one is limited largely by the inherent inaccuracy of the PY model, low bond dimension TT approximation are a good way to speed up the calculation.

\section{Discussion}

When constructing a tensor network representation of a physical problem, the first objective is identifying what aspect of the function is most compressible. In our two cases we took advantage of the weak correlations between dependencies on each spatial dimension, $(x,y,z)$ in the standard TT format, or the weak correlations between spatial scales as in the QTT format. Additionally, a source of high-dimensionality that does not seem easily compressible is that of high-order correlation functions. In this work on two-particle correlation functions, the two particles were lumped into single cores in both TT and QTT formats (e.g. Fig.~\ref{fig:QTT_diagram}). For example, for a one-dimensional problem a $p^{th}$ order correlation function would be highly difficult to compress in either format as it would either require very large bond dimensions or very large cores. Extending this work to higher order correlation functions thus would require rethinking the possible representation. Going beyond tensor trains to more general tensor networks, such as projected entangled pair states, may allow one to take advantage of both possible sources of compression, but one has to deal with the computational cost and difficulty of tensor diagrams with loops. 

Yukawa plasmas and other screened plasma pair-potentials are another major application of interest, for example the field of dusty plasmas \cite{pan2021stronglycoupledyukawaplasma}, and the switch from LJ potentials to Yukawa pair potentials requires a switch from a PY closure to the HNC closure for reasonable accuracy. This requires an exponential of a tensor train, which is generally possible, but requires the solution of a differential equation \cite{Eigel_2023}, or an expensive interpolation procedure involving TCI for QTT's \cite{Ritter_2024} which we have found to be slower and unstable compared to the PY methods.

An alternative to the standard self-consistent iteration algorithms employed in this paper are core-based optimization techniques based on density matrix renormalization group (DMRG) ideas \cite{white1993density} such as \cite{dolgov2014alternating, oseledets2011dmrg}. We did not find a local or core-based optimization scheme competitive with the standard all-core, global, mixing techniques we describe above. Doing local optimization naively requires constructing very large environment tensors due to the many non-linear (but simple) addition and multiplication operations required to obtain the self-consistent equation $h_{i+1}=A(h_i)$ where $A$ is the high-order non-linear operator that obtains a new $h$ based on the OZ and PY equations.   

Another approach to dimensionality reduction for such equilibrium problems is the coarse-grained approach to many-body theory \cite{noid2023perspective}, intuitively related to the scale-separation ideas QTTs rely on. In these problems and generically for the study of soft matter, non-spherically symmetric potentials are needed. In this paper, a spherically symmetric pair potential was assumed, however the original PY, HNC, and OZ equations were completely general initially, and our methods can be easily generalized, although with greater computational expense. Connecting the scale dependence of coarse-grained mappings \cite{kidder2024analysis} would be of significant theoretical and practical interest.

%%%%%%%%%%%%%%%%%%%%%%%%%%%%%%%%%%%%%%%%%%%%%%%%%%%%
%%%%%%%%%%%%%%%%%%%   SECTION  %%%%%%%%%%%%%%%%%%%%%
%%%%%%%%%%%%%%%%%%%%%%%%%%%%%%%%%%%%%%%%%%%%%%%%%%%%

\section{Conclusions}
Tensor trains and quantized tensor trains offer a general mechanism for compressing pair correlation functions and are efficient representations for solving the integral equations describing homogeneous or confined plasmas and liquids. We consider tensor trains with cores representing spatial dimensions and solve for the one and two particle density correlations for a liquid between parallel plates and show that this leads to a more efficient solution than the standard grid counterpart. We also considered the homogeneous problem in a quantized tensor train format where the cores represent different spatial scales, demonstrating high compression scaling as the log of the resolution.    

\section{Data Availability}
Data supporting the findings of this study are
available from the corresponding author on a reasonable
request.

\begin{acknowledgments}
We thank Chris Gerlach for improving our understanding of the literature in the early stages of this project. This work was supported by the Lawrence Livermore National Laboratory through the Academic Collaboration Team: University Partnerships (ACT UP) program. This work performed under the auspices of the U.S. Department of Energy by Lawrence Livermore National Laboratory under Contract LLNL-B659440 with document ID LLNL-JRNL-2008717. Additionally, this work was supported in part through computational resources and services provided by the Institute for Cyber-Enabled Research at Michigan State University.
\end{acknowledgments}

\section*{Author Contributions}
All authors contributed equally to conceptualization and editing. ZJ wrote the original draft, developed the methodology, and performed the formal analysis and investigation. MM and PG funded and supervised the project.

%%%%%%%%%%%%%%%%%%%%%%%%%%%%%%%%%%%%%%%%%%%%%%%%%%%%
%%%%%%%%%%%%%%%%%%%     BIB    %%%%%%%%%%%%%%%%%%%%%
%%%%%%%%%%%%%%%%%%%%%%%%%%%%%%%%%%%%%%%%%%%%%%%%%%%%

\bibliography{bib}
\appendix

%%%%%%%%%%%%%%%%%%%%%%%%%%%%%%%%%%%%%%%%%%%%%%%%%%%%
%%%%%%%%%%%%%%%%%%%  APPENDIX  %%%%%%%%%%%%%%%%%%%%%
%%%%%%%%%%%%%%%%%%%%%%%%%%%%%%%%%%%%%%%%%%%%%%%%%%%%

\section{Fourier Transforming Quantized Tensor Trains}
In these sections we seek to find the 3D FT of functions with either cylindrical or spherical symmetry.
\begin{align}
    \tilde{f}(\vb k) = \int d^3 \vb r e^{-i \vb k \cdot \vb r } f(\vb r)
\end{align}

%%%%%%%%%%%%%%%%%%% SUBSECTION %%%%%%%%%%%%%%%%%%%%%

\subsection{Hankel Transform for Cylindrical Symmetry Geometries} \label{app:FT_Hankel}
The Hankel transform is obtained in the case of an FT of a function that is cylindrically symmetric around one axis, where the FT is done only in the two-dimensional transverse space. 
\begin{align}
    f(x_1, x_2, \vec{k}) 
    &= \int d^2 \vec{\rho}_{12} f(x_1, x_2, \vec{\rho}_{12}) e^{i \vec{k} \cdot \vec{\rho}_{12} }  \nonumber \\
    f(x_1, x_2, k) 
    &= 2 \pi \int d \rho \rho  f(x_1, x_2, \rho) J_0 (k r),
\end{align}
where we have used $2 \pi J_0(x) = \int_0^{2\pi} e^{ix \cos \theta}$. The inverse FFT is 
\begin{align}
    f(x_1, x_2, \vec{\rho}_{12}) &= \int \frac{d^2 \vec{k}}{ (2 \pi)^2} f(x_1, x_2, \vec{k}) e^{-i \vec{k} \cdot \vec{\rho}_{12} } \nonumber \\
     f(x_1, x_2, \rho)  &= \frac{1}{2 \pi} \int d k k f(x_1, x_2, k) J_0 (k r).
\end{align}
We apply this FT to the case in Sec.~\ref{sec:TT_1D_asymmetry}, which involves a TT rather than a QTT. In this case, we can FT only the middle dimension of the one core. We use the package \cite{PyHank} to do this which optimizes the transform accuracy by optimizing grid placement. 

%%%%%%%%%%%%%%%%%%% SUBSECTION %%%%%%%%%%%%%%%%%%%%%

\subsection{Discrete Sine Transform for Spherically Symmetry Geometries} \label{app:FT_dst}
The FT can be represented as a single operator or tensor train matrix, that transforms the QTT into Fourier space, however this operator is incompressible by itself, but thankfully an alternative exists \cite{Dolgov2012}.  
First, we must transform our problem into a standard Fourier transform in one-dimension. Starting in three-dimensions, including spherical symmetry turns the transform into a sine transform as,
\begin{align}
    \tilde{f}(k) &= \frac{4 \pi}{k} \int_0^\infty dr r f(r) \sin(kr)\\
                 &= \frac{4 \pi}{k} \mathcal{F}_S[g](k) 
\end{align}
Where $\mathcal{F}_S$ represents the sine transform of $g(r)=r f(r)$. In particular, we use the type-4 discrete sine transform, 
\begin{align}
    \tilde{g}_j/\Delta r &= \sum_{i=0}^{N-1} f_i \sin(x_i k_j).
\end{align}
Where our discretized grid of length $N$ over the range $x \in (0,R)$ is defined with $x_{i} = \Delta x (i+1/2)$,  $k_i=\Delta k (i + 1/2)$ for $\Delta x = R/N$, $\Delta k = \pi/R$. This is equivalent to assuming $g(r)$ (f(r)) is odd (even) around $x=0$ and even around $x=R$ with a period of $2R$. We then reframe this as a discrete Fourier transform via
\begin{align}
    \tilde{g}_j/\Delta r &= -\Im\left[ \sum_{i=0}^{N-1} f_i \exp[-i\frac{\pi}{N} (i+1/2)(j+1/2)]\right],\nonumber \\
    \tilde{g}_j/\Delta r &= -\Im\left[e^{-i\frac{\pi}{2N} (j+1/2)} \sum_{i=0}^{N-1} f_i e^{-i\frac{\pi}{N} i(j+1/2)}\right],\nonumber\\
    \tilde{g}_j/\Delta r &= -\Im\left[e^{-i\frac{\pi}{2N} (j+1/2)} \sum_{i=0}^{N-1} \hat{f}_i e^{-i\frac{\pi}{N} i j}\right],\nonumber\\
    \tilde{g}_j/\Delta r &= -\Im\left[e^{-i\frac{\pi}{2N} (j+1/2)} \sum_{i=0}^{M-1} \hat{f}_i e^{-2i\frac{\pi}{M} i j}\right].
\end{align}
Here we define $\hat{f}_i=f_i e^{-i\frac{\pi i}{2N}}$ which we note is straightforward since exponentials with linear arguments have bond dimension one. Note in the last line we extend the domain by a factor of two and use $M=2N$, with $f_{i > N}=0$ to finally obtain the form of a standard Fourier transform. Extending the domain is extremely cheap with QTT's as it corresponds to adding one core only. To do the discrete Fourier transform in the last line, we use the ttpy package \cite{ttpy} which applies the superfast FT \cite{Dolgov2012}. 

%%%%%%%%%%%%%%%%%%%%%%%%%%%%%%%%%%%%%%%%%%%%%%%%%%%%
%%%%%%%%%%%%%%%%%%%   SECTION  %%%%%%%%%%%%%%%%%%%%%
%%%%%%%%%%%%%%%%%%%%%%%%%%%%%%%%%%%%%%%%%%%%%%%%%%%%

\section{Geometrical Symmetries and Degrees of Freedom} \label{app:dof}
In an external potential that depends on $a$ space coordinates, the number of degrees of freedom of an $p-$particle function $f^{(p)}(\r_1, \cdots \r_p)$ in $d$ dimensions is,
\begin{align}
\label{eq:dof}
    DOF(p, a, d) = pd - \sum_{j=a}^{p-1} {\rm max}(0, d-j) 
\end{align}
The two problems we consider in this paper are for $p=1, 2$ and in Sec.~\ref{sec:QTT_Homo}, $a=0$ and in Sec.~\ref{sec:TT_1D_asymmetry}, $a=1$.  
 
\begin{table*}[h]
\centering
\begin{tabular}{|c|c|c|c|c|c|c|c|c|c|l}
\hline
\multicolumn{2}{|c|}{} & \multicolumn{2}{c|}{$(a=0)$ Homogeneous } & \multicolumn{2}{c|}{$\phi(z) (a=1)$, e.g. Parallel walls} & \multicolumn{2}{c|}{ $\phi(x,y) (a=2)$, e.g. 2D box} & \multicolumn{2}{c|}{ $\phi(x,y,z) (a=3)$, Fully Heterogeneous} \\ \hline
\textbf{p} & \textbf{DOF} & \textbf{Sym.} & \textbf{DOF-Sym.}     & \textbf{Sym.} &\textbf{DOF-Sym.}     & \textbf{Sym.} &\textbf{DOF-Sym.}     & \textbf{Sym.} & \textbf{DOF-Sym.}   \\ \hline
1          &      3        &     3          &    0                &  2          &    1                   &  1          &    2                   & 0          &    3                  \\ \hline
2          &      6        &     2          &    1                &  1          &    3                   &  0          &    5                   & 0          &    6                  \\ \hline
3          &      9       &      1         &     3                &  0         &     6                   &  0         &     8                   & 0          &    9                  \\ \hline
\vdots     & \vdots       & \vdots        &        \vdots         &   \vdots  &    \vdots                &\vdots     & \vdots                   & \vdots          &  \vdots                      \\ \hline
p          &      3p        &   0            &  3p-6              & 0            &  3p-3                 & 0            &  3p-1                 & 0          &   12                  \\ \hline
\end{tabular}
\caption{The degrees of freedom of some n-particle correlation function $f^{(n)}(\r_1, \cdots \r_n)$ in an $a$ dimensional external potential in 3 dimensions. Note that the degree of asymmetry changes substantially the degrees of freedom needed only for small n.}
\label{tab:your_label}
\end{table*}

% \section{Methods Summary}\label{app:algorithms}
% The specific order that one implements a set of self-consistent equations can modify the convergence properties, so here we state exactly the algorithm we used.

% \begin{algorithm}[H]
% \caption{Self-consistent solution of PY/OZ equations in QTT format}
% \label{alg:py_oz_qtt}
% \begin{algorithmic}[1]
% \STATE Initialize $n$, $h$, and $c$ using TT-SVD for TTs, and TCI for QTTs.
% \REPEAT
%     \STATE Solve Ornstein-Zernike (OZ) equation for $\gamma(\mathbf{r})$:
%     \[
%         \gamma(\mathbf{r}) = c(\mathbf{r}) + n(\mathbf{r}) \int c(\mathbf{r}-\mathbf{r}') h(\mathbf{r}') d\mathbf{r}'
%     \]
%     using efficient QTT convolution
%     \STATE Update $h(\mathbf{r})$ from the PY closure:
%     \[
%         h(\mathbf{r}) = f(\mathbf{r}) + [1 + f(\mathbf{r})] \gamma(\mathbf{r})
%     \]
%     \STATE Update $c(\mathbf{r})$:
%     \[
%         c(\mathbf{r}) = f(\mathbf{r}) [1 + \gamma(\mathbf{r})]
%     \]
%     \STATE Update density $n(\mathbf{r})$ using a discretized Lovett–Mou–Buff equation (integrated by parts)
%     \STATE Apply linear mixing to stabilize updates:
%     \[
%         u^{(k+1)} = \alpha u_{\text{new}} + (1-\alpha) u^{(k)}
%     \]
%     where $u \in \{n, h, c\}$ and $\alpha$ is the mixing parameter
%     \STATE Round QTT ranks to tolerance $\epsilon$
% \UNTIL Convergence: relative residuals of $n$, $h$, and $c$ below specified tolerance
% \end{algorithmic}
% \end{algorithm}

\end{document}